\documentclass[twocolumn,floatfix,superscriptaddress,showpacs,preprintnumbers,prl]{revtex4}
\usepackage{graphicx}
\usepackage{epsfig}
\usepackage{amssymb}

\newcommand{\INOA}{CNR- Istituto dei Sistemi Complessi,
Via Madonna del Piano, 10, 50019 Sesto Fiorentino (FI), Italy \\
and the Italian Embassy in Tel Aviv, Trade Tower, 25 Hamered
Street, Tel Aviv, Israel}
\newcommand{\NIZ}{Department of Radiophysics, Nizhny Novgorod University, 23,
Gagarin Avenue, 603600 Nizhny Novgorod, Russia }
\newcommand{\CAT}{Dipartimento di Fisica e Astronomia,
Universit\`a di Catania, and INFN Sezione di Catania, Via S.
Sofia, 64, 95123 Catania, Italy}

\begin{document}
\title{Detection of Complex Networks Modularity by Dynamical Clustering}

\author{S. Boccaletti}\affiliation{\INOA}
\author{M. Ivanchenko}\affiliation{\NIZ}
\author{V. Latora}\affiliation{\CAT}
\author{A. Pluchino}\affiliation{\CAT}
\author{A. Rapisarda}\affiliation{\CAT}


\date{29 March 2007 -
PRE  Rapid  Communications in press}
\begin{abstract}

Based on cluster de-synchronization properties of phase
oscillators, we introduce an efficient method for the detection
and identification of modules in complex networks. The performance
of the algorithm is tested on computer generated and real-world
networks whose modular structure is already known or has been
studied by means of other methods. The algorithm attains a high
level of precision, especially when the modular units are very
mixed and hardly detectable by the other methods, with a
computational effort ${\cal O}(KN)$ on a generic graph with $N$
nodes and $K$ links.
\end{abstract}
\maketitle

Hierarchical modular structures constitute one of the most
important property of real-world networked systems \cite{report}.
For instance, tightly connected groups of nodes in a social
network represent individuals belonging to social communities,
while modules in cellular and genetic networks are somehow related
to functional modules. Consequently, identifying the modular
structure of a complex network is a crucial issue in the analysis
and understanding of the growth mechanisms and the processes
running on top of such networks. {\it Modules} (called sometimes
{\it community structures} in social science) are tightly
connected subgraphs of a network, i.e. subsets of nodes {\it
within} which the network connections are dense, and {\it between}
which connections are sparser. Nodes, indeed, belonging to such
tight-knit modules, constitute units that separately contribute to
the collective functioning of the network. For instance, subgroups
in {\it social networks} often have their own norms, orientations
and subcultures, sometimes running counter to the official
culture, and are the most important source of a person's identity.
Analogously, the presence of subgroups in {\it biological} and
{\it technological} networks is at the basis of their functioning.

\noindent The detection of the modular structure of a network is
formally equivalent to the classical {\it graph partitioning}
problem in computer science, that finds many practical
applications such as load balancing in parallel computation,
circuit partitioning and telephone network design, and is known to
be a NP-complete problem \cite{gj79}. Therefore, although modules
detection in large graphs is potentially very relevant and useful,
so far this trial has been seriously hampered by the large
associated computational demand. To overcome this limitation, a
series of efficient heuristic methods has been proposed over the
years. Examples include methods based on {\it spectral analysis}
\cite{psl90}, or the {\it hierarchical clustering methods}
developed in the context of social networks analysis
\cite{wasserman94}, or methods allowing for multi-community
membership \cite{vicsek}. More recently, Girvan and Newman (GN)
have proposed an algorithm which works quite well for general
cases \cite{ng04}. The GN is an iterative divisive method based in
finding and removing progressively the edges with the largest
betweenness, until the network breaks up into components. The
betweenness $b_{ij}$ of an edge connecting nodes $i$ and $j$ is
defined as the number of shortest paths between pairs of nodes
that run through that edge \cite{ng04}. As the few edges lying
between modules are expected to be those with the highest
betweenness, by removing them recursively a separation of the
network into its communities can be found. Therefore, the GN
algorithm produces a hierarchy of subdivisions of a network of $N$
nodes, from a single component to $N$ isolated nodes. In order to
know which of the divisions is the best one, Girvan and Newman
have proposed to look at the maximum of the modularity $Q$, a
quantity measuring the degree of correlation between the
probability of having an edge joining two sites and the fact that
the sites belong to the same modular unit (see Ref.~\cite{ng04}
for the mathematical definition of $Q$). The GN algorithm runs in
${\cal O}(K^2 N)$ time on an arbitrary graph with $K$ edges and
$N$ vertices, or ${\cal O}(N^3)$ time on a sparse graph. In fact,
calculating the betweenness for all edges requires a time of the
order of $KN$ \cite{bet}, since it corresponds to the evaluation
of all-shortest-paths (ASP) problem. And the betweenness for all
edges need to be recalculated every time after an edge has been
removed (the betweenness recalculation is a fundamental aspect of
the GN algorithm) \cite{ng04}. This restricts the applications to
networks of at most a few thousands of vertices with current
hardware facilities. Successively, a series of faster methods
directly based on the optimization of $Q$ have been proposed
\cite{optimization,various}, which allow up to a ${\cal O}(N \log^2 N)$
running time for finding modules in sparse graphs.

All the above mentioned methods are based on the structure of a
network, meaning that they use solely the information contained in
the adjacency matrix $A=\{a_{ij}\}$ (or any equivalent
representation of the topology) of the graph. As complementary to
such approaches, the authors of Ref.~\cite{rb04} have proposed a
method to find modules based on the {\em statistical properties}
of a system of spins (namely q-state Potts spins) associated to
the nodes of the graphs. In this Letter we propose a {\it
dynamical clustering} (DC) method based on the properties of a
dynamical system associated to the graph. DC techniques were
initiated by the relevant observation that topological hierarchies
can be associated to dynamical time scales in the transient of a
synchronization process of coupled oscillators \cite{arenas}.
Although being fast, so far DC methods do not provide in general
the same accuracy in the identification of the communities.

Here, we show how to combine topological and dynamical information
in order to devise a DC algorithm that is able to identify the
modular structure of a graph with a precision competitive with the
best techniques based solely on the topology. The method we
present is based upon the well-known phenomenon of synchronization
clusters of non identical phase oscillators \cite{boc02}, each one
associated to a node, and interacting through the edges of the
graph. Clusters of synchronized oscillators represent an
intermediate regime between global phase locking and full absence
of synchronization, thus implying a division of the whole graphs
into groups of elements which oscillate at the same (average)
frequency. The key idea is that, starting from a fully
synchronized state of the network, a dynamical change in the
weights of the interactions, that retain information on the
original betweenness distribution, yields a progressive
hierarchical clustering that fully detects modular structures.

For the sake of illustration and without lack of generality, we
specify our technique with reference to the so called {\it Opinion
Changing Rate} (OCR) model, a continuous-time system of coupled
phase oscillators introduced for the modeling of opinion consensus
in social networks  \cite{ocr}, and representing a variation of
the Kuramoto model \cite{kura}. Other continuos-time (Kuramoto and
R\"{o}ssler dynamics), and also discrete-time (coupled circle
maps) dynamical systems have been investigated and will be
reported elsewhere. Given a undirected, unweighted graph with $N$
nodes and $K$ edges, described by the adjacency matrix
$A=\{a_{ij}\}$, we associate to each node $i$ ($i=1,\dots,N$) a
dynamical variable $x_i(t) \in ] -\infty, + \infty [$. The
dynamics of each node is governed by:
\begin{equation}
    \dot{x_i} (t)  = \omega_i + \frac{\sigma}{\sum_{j\in {\cal N}_i} b_{ij}^{\alpha(t)}}\sum_{j\in {\cal N}_i}
      ~ b_{ij}^{\alpha(t)}  \sin (x_j  - x_i) \beta e^{- \beta |x_j  - x_i|
      }
\label{eq:OCR}
\end{equation}
where $\omega_i$ is the natural frequency of node $i$ (in the
numerical simulations the $\omega_{i}\text{'s}$ are randomly
sorted from a uniform distribution between $\omega_{\text min}=0$
and $\omega_{\text max}=1$), $\sigma$ is the coupling strength,
and ${\cal N}_i$ is the set of nodes adjacent to $i$, i.e. all
nodes $j$ for which $a_{ij}=a_{ji}=1$. The constant parameter
$\beta$, tuning the exponential factor in the coupling term of
Eqs. (\ref{eq:OCR}), switches off the interaction when the phase
distance between two oscillators exceeds a certain threshold (as
usual \cite{ocr} we fix $\beta=3$). Notice that the interaction
between two adjacent nodes $i$ and $j$ is weighted by the term $
b_{ij}^{\alpha(t)} / \sum_{j\in {\cal N}_i}  b_{ij}^{\alpha(t)}$,
where $b_{ij}$ is the betweenness of the edge $i,j$, and
$\alpha(t)$ is a time dependent exponent, such that $\alpha(0)=0$.
In Ref.\cite{SPRL} it has been shown that the ability of a
dynamical network, as the one in Eqs. (\ref{eq:OCR}), to maintain
a synchronization state crucially depends on the value of the
parameter $\alpha$. For such a reason, in the DC algorithm to find
modular structures, we fix the coupling strength $\sigma$ equal to
an arbitrary value such that the unweighted ($\alpha=0$) network
is fully synchronized, and we solve Eqs. (\ref{eq:OCR}) with a
progressively (stepwise) decreasing value of $\alpha(t)$. Namely,
while keeping fixed $\sigma$, we consider $\alpha(t_{l+1})\alpha(t_l)- \delta \alpha$ for $t_{l+1}> t > t_l$, where
$t_{l+1}-t_l=T ~\forall l$ (in the following $T=2$), and  $\delta
\alpha$ is a parameter that will be specified below. As the edges
connecting nodes belonging to the same module (to two different
modules) have in general small (large) betweenness, when $\alpha$
decreases from zero, the corresponding interaction strengths on
those edges become increasingly enhanced (weakened). Since the
network is prepared to be fully synchronized, it has to be
expected that, as $\alpha$ decreases, the original synchronization
state hierarchically splits into clusters of synchronized
elements, accordingly to the hierarchy of modules present in the
graph. The individuation of synchronization clusters is made in
terms of groups of nodes with the same instantaneous frequency
$\dot{x}(t)$. The procedure consists then in monitoring the
emerging set of synchronization clusters at each value of
$\alpha(t)$. The best division in communities of the graph (the
best $\alpha$ value) is individuated by looking at the maximum (as
a function of $\alpha$) of the modularity $Q$ \cite{ng04}.

\begin{figure}
\begin{center}
\epsfig{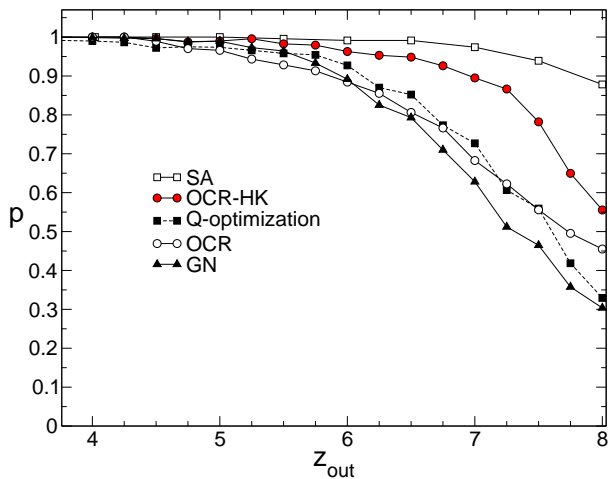}
\end{center}
\caption{Fraction $p$ of correctly identified nodes as a function
of $z_{out}$ (average number of inter-modular edges per node) for
computer generated graphs with $N=128$ nodes, four communities and
an average degree $<k>=16$. The results of DC methods based
respectively on the OCR (open circles) and the OCR-HK (full
circles) models, are compared with three standard methods, two
based solely on the topology, such as the GN algorithm (full triangles)
\cite{ng04} and the Newman Q-optimization fast algorithm
(full squares) \cite{optimization}, and one based on an evolutionary method,
the Simulated Annealing algorithm \cite{danon} (open squares).} \label{test}
\end{figure}

%
In order to comparatively evaluate the performance of the
algorithm, we have considered, as in Ref.~\cite{ng04}, a set of
computer generated random graphs constructed in such a way to have
a well defined modular structure. All graphs are generated with $N
= 128$ nodes and $K=1024$ edges. The nodes are divided into four
communities, containing 32 nodes each. Pairs of nodes belonging to
the same module (to different modules) are linked with probability
$p_{in}$ ($p_{out}$). $p_{out}$ is taken so that the average
number $z_{out}$ of edges a node forms with members of other
communities can be controlled. While $z_{out}$ can be then varied,
$p_{in}$ is chosen so as to maintain a constant total average node
degree $<k>=16$. As $z_{out}$ increases, the modular structure of
the network becomes therefore weaker and harder to identify. As
the {\it real} modular structure is here directly imposed by the
generation process, the accuracy of the identification method can
be assessed by monitoring the fraction $p$ of correctly classified
nodes vs. $z_{out}$. In Fig.~\ref{test} we report the value of $p$
(averaged over twenty different realizations of the computer
generated graphs and of the initial conditions) as a function of
$z_{out}$, for the DC algorithm based on the OCR model of Eqs.
(\ref{eq:OCR}), with $\sigma=5.0$ and $\delta\alpha=0.1$. The
resulting performance (open circles) is comparable to that of the
best methods based solely on the topology, such as the GN (full
triangles) \cite{ng04} and the Newman Q-optimization fast algorithm (full squares)
\cite{optimization}.
%

%
The performance of the DC algorithm considered can be made better
by adding a simple modification to the OCR model which further
stabilizes the system. Such modification consists in changing in
time the natural frequencies $\omega$'s according to the idea of
{\em confidence bound} introduced for the first time by Hegselmann
and Krause (HK), in the context of models for opinion formation
\cite{heg}. Therefore, we will refer to the improved method as the
$OCR-HK$ dynamical clustering. The confidence bound is a parameter
$\epsilon$ which fixes the range of compatibility of the nodes. At
each time step, the generic node $i$, having a dynamical variable
$x_i(t)$ and a natural frequency $\omega_i(t)$, checks how many of
its neighbors $j$ are compatible, i.e. have a value of the
variable $x_j(t)$ falling inside the confidence range
[$x_i-\epsilon,x_i+\epsilon$]. Then, at the following step in the
numerical integration, we set $\omega_i(t+ \Delta t)$, i.e. the
node takes the average value of the $\omega$'s of its compatible
neighbors at time $t$. In the $OCR-HK$, the changes of the
$\omega(t)$'s is superimposed to the main dynamical evolution of
Eq.(\ref{eq:OCR}) and noticeably contributes to stabilize the
frequencies of the oscillators according to the correct modular
structure of the network, also reducing the dependence of the
algorithm on the initial conditions. The results with computer
generated graphs ($\delta\alpha=0.1$) are reported in
Fig.~\ref{test} as full circles. The improvement in the
performance of the OCR-HK method with respect to both the standard
OCR and the two topological methods (GN and Q-optimization), is
evident for $z_{out}> 5$, and it can be made even larger using for
smaller values of $\delta\alpha$. For completeness, we also report
the results of an optimization procedure based on a Simulated
Annealing (SA) \cite{danon}, which is presently the most accurate
method available on the market, though very CPU time consuming. As
for the computational cost, our algorithm provides an improvement
with respect to the majority of other methods (see Table 1 in
Ref.\cite{danon}). For instance, while iterative topological
algorithms \cite{ng04,various} need to recalculate the betweenness
distribution all the times a given edge is removed, in our case
that distribution  has to be evaluated {\it only} for the initial
graph, as the cluster de-synchronization process itself gives a
progressive weakening of the edges with highest betweenness. We
have analyzed sparse graphs of size up to N=16,384 and  found a
scaling law of ${\cal O}(N^{1.76})$ for the dynamical evolution of
the OCR-HK system. However, since the initial calculation of
betweenness takes ${\cal O}(N^2)$ operations, the total CPU time
scales as ${\cal O}(N^2)$ too. Considering that the fastest
algorithm on the market (${\cal O}(N \log^2 N)$) is the
Q-optimization one \cite{optimization}, which on the other hand is
less accurate than the OCR-HK (as shown in Fig.~\ref{test}), we
can conclude that our DC method provides an excellent compromise
between  accuracy and computational demand \cite{danon}.

It should be noticed that the proposed method conceptually differs
from the dynamical clustering technique introduced in Ref.
\cite{arenas}. Indeed, while in \cite{arenas} the modular
hierarchy of a network was associated to different time scales in
the transient dynamics toward a fully synchronized dynamics, here
it corresponds to a hierarchical sequence of {\it asymptotically}
synchronized states, from the initial ($\alpha(0)=0$) full network
synchronization, to progressive cluster synchronization as
$\alpha(t)$ decreases. A relevant consequence is that our
technique is almost fully insensitive to differences in the
initial conditions for the phases of the coupled oscillators, as
far as the local dynamics is selected to have a unique attractor.

Finally, we tested how the method works on two typical real-world
networks: the Zachary Karate Club network ($N=34$, $K=78$)
\cite{zachary} and the food web of marine organisms living in the
Chesapeake Bay ($N=33$, $K=71$) \cite{baird,eff}. In both cases we
have some a-priori knowledge of the existing modules. In fact, the
karate club network, is known to split into two smaller
communities, whose detailed composition was reported by Zachary
\cite{zachary}. Analogously, the food web organisms contain a main
separation in two large communities, according to the distinction
between pelagic organisms (living near the surface or at middle
depths) and benthic organisms (living near the bottom).
As in the previous simulations, we first calculate the set of edge
betweenness $\{ b_{ij} \}$ and then we integrate numerically
Eqs.(\ref{eq:OCR}) with the $HK$ modification on the $\omega$'s,
with $\delta\alpha=0.1$ and $\sigma=5.0$ (the latter ensures again
an initial fully synchronized state at $\alpha=0$).

\begin{figure}
\begin{center}
\epsfig{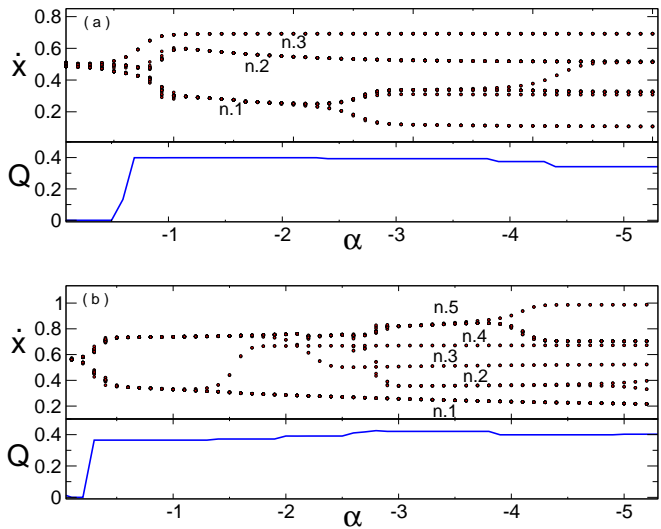}
\end{center}
\caption{ (Color online).
The distribution of instantaneous frequencies and the correspondent
modularity $Q$ are reported as a function of $\alpha$ for the OCR-HK model.
The application to the Karate Club network and to the Chesapeake Bay food web
are shown in panel (a) and in panel (b) respectively.
In both the simulations $\sigma=5.0$, $\delta\alpha=0.1$ and $\epsilon=0.0005$.}
\label{real}
\end{figure}

In Fig.~\ref{real} the $N$ instantaneous frequencies $\dot{x}_i$,
and the modularity $Q$, are plotted as a function of $\alpha$
(i.e. as a function of time) for both the karate club, panel (a),
and the food web network, panel (b).
In panel (a) the best configuration, with $Q\sim0.40$, is reached
around $-1.0\gtrsim\alpha\gtrsim-2.5$ and yields a partition of
the karate club network into three stable communities which very
well describe the real situation. The largest one, labelled with n.1
in the figure (nodes $9,10,29,31,15,16,19,21,23,32,33,34,24,25,26,27,28,$ $30$),
fully corresponds to one of the two communities reported by Zachary,
while the sum of the remaining two communities, labelled as n.2
(nodes $1,2,3,4,8,12,13,14,18,20,22$) and n.3 (nodes $5,6,7,11,17$),
corresponds to the second Zachary's module of $16$ elements. Notice that cluster
n.3 represents a very well connected subset
that is frequently recognized as a separated module also by other
methods \cite{ng04}.
Moreover, the value of the best modularity found is larger than
that of the Zachary partition into two communities ($Q\sim0.37$).
Analogously good performance is obtained for the food web.
In panel (b) the highest value of $Q$, namely $Q\sim0.42$,
is reached for $-2.8\gtrsim\alpha\gtrsim-3.8$, yielding
a division of the food web into five communities
(n.1:nodes $3,14,15,16,18,19,25,26,27,28,29$;
n.2:nodes $22,30,31,32$; n.3:nodes $5,6$; n.4:nodes $4,17$;
n.5:nodes $8,9,10,20,24,1,2,7,11,12,13,21,23,33$)
in which, with respect to Refs.\cite{ng04,eff},
the distinction between pelagic and benthic organisms is not only preserved
but also improved.

In conclusion, we have introduced an efficient algorithm for the
detection and identification of modular structures, that attains a
very high precision, with a small associated computational effort
that scales as ${\cal O} (N^2)$. Our method, therefore, could be
of use for a reliable modules detection in sizable networks (e.g.
biological, neural networks),  and can contribute to a better
understanding of the hierarchical functioning of networked systems
in many physical, biological and technological cases.

The Authors are indebted with A. Amann, F.T. Arecchi, M. Chavez,
R. L\'opez-Ruiz and Y. Moreno for the many helpful discussions on
the subject.  M.I. acknowledges projects n. RFBR 05-02-19815-MF-a,
06-02-16499-\'A, 06-02-16596-\'A. V.L., A.P. and A.R. acknowledge
financial support from the PRIN05-MIUR project "Dynamics and
Thermodynamics of Systems with Long-Range Interactions".

\end{document}